\documentclass[sigconf, nonacm]{acmart}

\newcommand\vldbdoi{10.14778/3611540.3611620}
\newcommand\vldbpages{4058 - 4061}
\newcommand\vldbvolume{16}
\newcommand\vldbissue{12}
\newcommand\vldbyear{2023}
\newcommand\vldbauthors{\authors}
\newcommand\vldbtitle{\shorttitle} 
\newcommand\vldbpagestyle{empty}

\usepackage{tabularx}

\newcommand{\system}[0]{\textsc{Cornet}}
\usepackage{tikz}
\newcommand*\circled[1]{\textcircled{\raisebox{-0.8pt}{#1}}}

\begin{document}

%%
%% The "title" command has an optional parameter,
%% allowing the author to define a "short title" to be used in page headers.
\title{\system{}: Learning Spreadsheet Formatting Rules By Example}

\author{Mukul Singh}
\affiliation{%
  \institution{Microsoft}
  \city{Delhi}
  \country{India}
  \postcode{43017-6221}
}
\email{singhmukul@microsoft.com}

\author{Jos\'e Cambronero Sanchez}
\affiliation{%
  \institution{Microsoft}
  \city{Redmond}
  \country{USA}
}
\email{jcambronero@microsoft.com}

\author{Sumit Gulwani}
\affiliation{%
  \institution{Microsoft}
  \city{Redmond}
  \country{USA}
}
\email{sumitg@microsoft.com}

\author{Vu Le}
\affiliation{%
  \institution{Microsoft}
  \city{Redmond}
  \country{USA}
}
\email{levu@microsoft.com}

\author{Carina Negreanu}
\affiliation{%
  \institution{Microsoft Research}
  \city{Cambridge}
  \country{UK}
}
\email{cnegreanu@microsoft.com}

\author{Gust Verbruggen}
\affiliation{%
  \institution{Microsoft}
  \city{Redmond}
  \country{USA}
}
\email{gverbruggen@microsoft.com}

%%
%% By default, the full list of authors will be used in the page
%% headers. Often, this list is too long, and will overlap
%% other information printed in the page headers. This command allows
%% the author to define a more concise list
%% of authors' names for this purpose.
\renewcommand{\shortauthors}{Trovato et al.}

%%
%% The abstract is a short summary of the work to be presented in the
%% article.
\begin{abstract}
Data management and analysis tasks are often carried out
using spreadsheet software. A popular feature in most spreadsheet
platforms is the ability to define data-dependent formatting rules.
These rules can express actions such as \emph{``color red all 
entries in a column that are negative''} or
\emph{``bold all rows not containing error or failure''}. Unfortunately, users who
want to exercise this functionality need to manually write these conditional
formatting (CF) rules. We introduce \system{}, a system that 
automatically learns such conditional formatting rules
from user examples. 
\system{} takes inspiration from inductive program synthesis and combines symbolic rule enumeration, based on  semi-supervised clustering and iterative decision tree learning, with a neural ranker to produce accurate conditional formatting rules.
% In this demonstration, we will introduce
% a simple interface 
% as an add-in to
% Microsoft's Excel, through which the user can provide
% one or two
% formatted cells as examples. \system{} can then learn 
% from these examples, generate a simple symbolic rule,
% and result in the user's desired formatting of their spreadsheet.
In this demonstration, we show \system{}
in action as a simple add-in to 
Microsoft's Excel. After the user provides one or two formatted cells as
examples, \system{} generates
formatting rule suggestions for the user to apply to the spreadsheet.

\end{abstract}

\maketitle

%%% do not modify the following VLDB block %%
%%% VLDB block start %%%
\pagestyle{\vldbpagestyle}
\begingroup\small\noindent\raggedright\textbf{PVLDB Reference Format:}\\
\vldbauthors. \vldbtitle. PVLDB, \vldbvolume(\vldbissue): \vldbpages, \vldbyear.\\
\href{https://doi.org/\vldbdoi}{doi:\vldbdoi}
\endgroup
\begingroup
\renewcommand\thefootnote{}\footnote{\noindent
This work is licensed under the Creative Commons BY-NC-ND 4.0 International License. Visit \url{https://creativecommons.org/licenses/by-nc-nd/4.0/} to view a copy of this license. For any use beyond those covered by this license, obtain permission by emailing \href{mailto:info@vldb.org}{info@vldb.org}. Copyright is held by the owner/author(s). Publication rights licensed to the VLDB Endowment. \\
\raggedright Proceedings of the VLDB Endowment, Vol. \vldbvolume, No. \vldbissue\ %
ISSN 2150-8097. \\
\href{https://doi.org/\vldbdoi}{doi:\vldbdoi} \\
}\addtocounter{footnote}{-1}\endgroup
%%% VLDB block end %%%

%%% do not modify the following VLDB block %%
%%% VLDB block start %%%
% \ifdefempty{\vldbavailabilityurl}{}{
% \vspace{.3cm}
% \begingroup\small\noindent\raggedright\textbf{PVLDB Artifact Availability:}\\
% The full paper can be accessed at \url{https://arxiv.org/abs/2208.06032}.
% \endgroup
% }
%%% VLDB block end %%%

\section{Introduction}

Millions of users~\cite{spreadsheet-usage} perform their data
management and analysis in 
spreadsheet software.
Most popular spreadsheet platforms,
such as Microsoft's Excel, allow users
to define data-dependent
formatting rules.
These rules can express actions such as \emph{``color red all entries in a column that are negative''} or \emph{``bold all rows not containing error or failure''}. 
These rules are typically called \emph{conditional formatting rules} (CF rules).
Unfortunately, users have to write conditional formatting rules manually, which requires both programming expertise and familiarity with the particular data platform and its rule language.

% Unfortunately, creating conditional formatting rules requires users to understand the syntax and logic behind them.
% As of March 2023, more than 10,000 conditional formatting related questions were posted on the Excel tech help community alone  \cite{ExcelHelpForum}.
A search on the Excel tech help community \cite{ExcelHelpForum} reveals more than 10,000 questions on conditional formatting as of March 2023.
We find that these posts tend to share
three key struggles, preventing
users from effectively using
conditional formatting.
First, many users are unaware of CF rules and instead manually format their spreadsheets.
Second, when a user does enter a CF rule, they can fail to do so correctly (because of invalid syntax or incorrect logic) and end up manually formatting the sheet.
Third, even when a user writes a valid rule that matches their desired formatting, these rules can be unnecessarily complex (and candidates for simplification) or not generalizable (may deviate from the desired behavior when data in the column changes or new data is added to the column).

\begin{figure}[t]
\centering
\includegraphics[width=0.85\columnwidth]{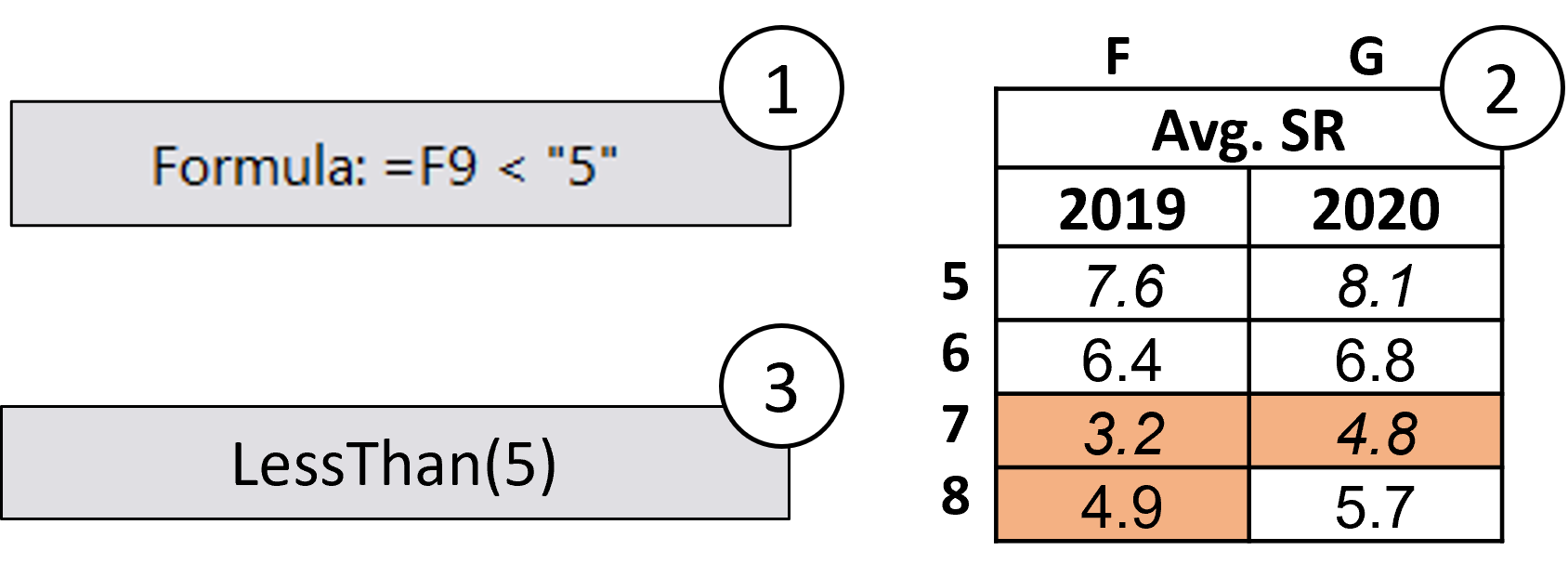}
\caption{Example of an incorrect
CF rule manually written by a user.
\circled{1} A user defined a custom rule that always evaluates to false (as it compares the cell contents to the string ``5'' instead of the numeric literal 5). 
% \jose{what is the relevance of "Rule (applied in order shown)"? If none, then remove}
\circled{2} Because the rule does not format any cell, the user had to manually format the sheet as reflected in this figure. \circled{3} The correct CF rule, which can be learned by \system{} from 2 examples.
% \jose{it is a bit confusing to reuse green color in multiple places, does it have meaning?}
} 
\label{fig:inccorect_cf}
\end{figure}

\begin{figure*}[t]
\centering
\includegraphics[width=\textwidth]{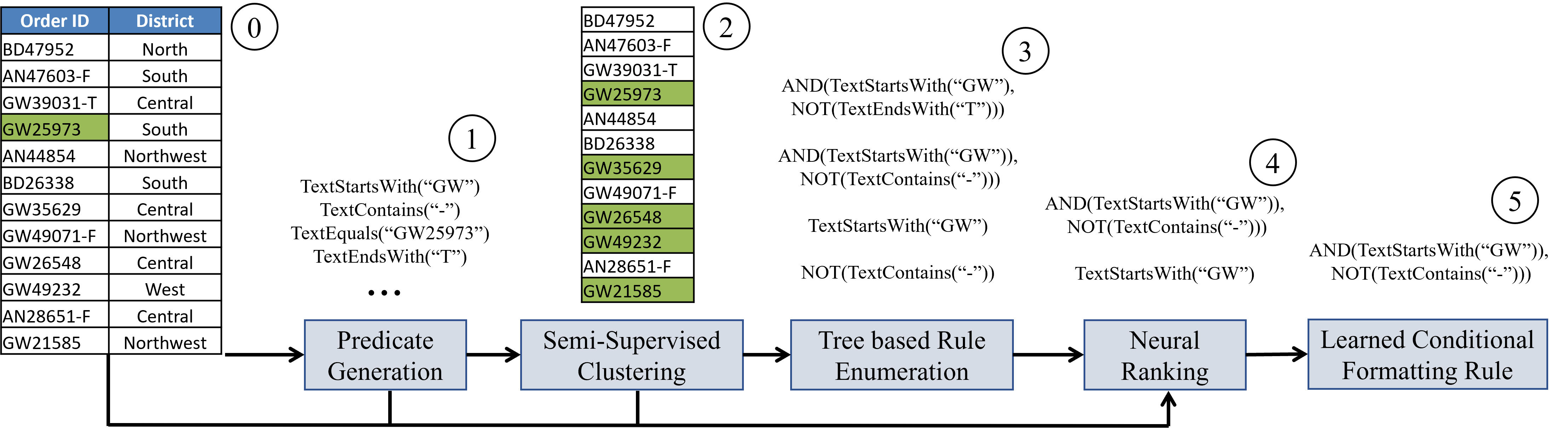} % Reduce the figure size so that it is slightly narrower than the column. Don't use precise values for figure width.This setup will avoid overfull boxes.
\caption{\system{}'s architecture illustrated through an example case: \circled{0} input table with a single user example (partial formatting), \circled{1} predicate generation for all cells in the table, \circled{2} semi-supervised clustering using examples and positional information to address the challenge of unlabeled cells, \circled{3} enumerating rules over the clustered column using multiple decision trees, \circled{4} neural ranker to score generated rules, and \circled{5} final learned conditional formatting rule.}
\label{architecture}
\end{figure*}

Some of these struggles are reinforced by the need to write custom formulas.
For example, in Excel CF rules that require logical operations like \textsc{or} and \textsc{and} must be written as custom formulas. Additionally, there is no validation for data types in the standard CF interface, causing surprising results. For example, a rule can (incorrectly) compare a text column to a numeric constant.

\begin{example}
Figure~\ref{fig:inccorect_cf} shows a public spreadsheet where the user wanted to highlight cells with value less than 5. Unfortunately, they wrote a rule that incorrectly uses the string ``5'' instead of the number. Ultimately, they manually formatted the sheet.
\end{example}

In this paper we present a demonstration of \system{}
\footnote{The full paper can be accessed at \url{https://arxiv.org/abs/2208.06032}.}
(\textbf{C}onditional \textbf{ORN}amentation by \textbf{E}xamples in \textbf{T}ables),
a system that allows users to automatically generate a conditional formatting rule from examples, thus mitigating the challenges previously outlined. \system{} takes a small number of user formatted cells as input to learn the most likely formatting rule that generalizes to other cells.
\system{} explores possible predicates for the target column, hypothesizes cell grouping via semi-supervised clustering and then learns candidate rules by employing an iterative tree learning procedure. Since multiple rules may result from this procedure, we train a neural ranker to return the most likely CF rule, given the spreadsheet data, user examples, and
candidate rules' execution on the data.

\begin{example}\label{arch-example}
    Consider the example shown in Figure~\ref{architecture}, the user wants to color all IDs that start with "GW" but don't contain "-F" or "-T" at the end. To accomplish this via the Microsoft Excel GUI, the user will have to navigate 2 dialog boxes, 3 dropdown menus, perform 9 clicks, and then write the Excel formula \texttt{AND(LEFT(A2,2)="GW", NOT(OR(RIGHT(A2,2)="-F", RIGHT(A2\\,2)="-T")))}. With \system{}, the user can simply color the fourth cell in Figure~\ref{architecture} \circled{0}, and \system{} learns and applies the correct rule.
\end{example}

In our demonstration, we will show an implementation of a lightweight interface in Microsoft Excel that exercises \system{} and can successfully learn the rule in Example~\ref{arch-example} and more complex scenarios. 
Our interface first allows users to provide one or two
examples of their desired formatting.
\footnote{Text rules can be learnt accurately by providing just one or two examples. Numeric columns typically require more examples because of their large search space.}
They can then invoke \system{}, which
will learn a symbolic rule that
covers the examples and generalizes to
the rest of the column.
Finally, \system{} applies the rule learned to format
the spreadsheet.
We show how both novice and experienced spreadsheet users can use \system{}.

In the remainder of this paper,
we present a high-level description of \system{}
and provide an outline of our proposed demonstration.
\begin{figure*}[t]
\centering
\includegraphics[width=\textwidth]{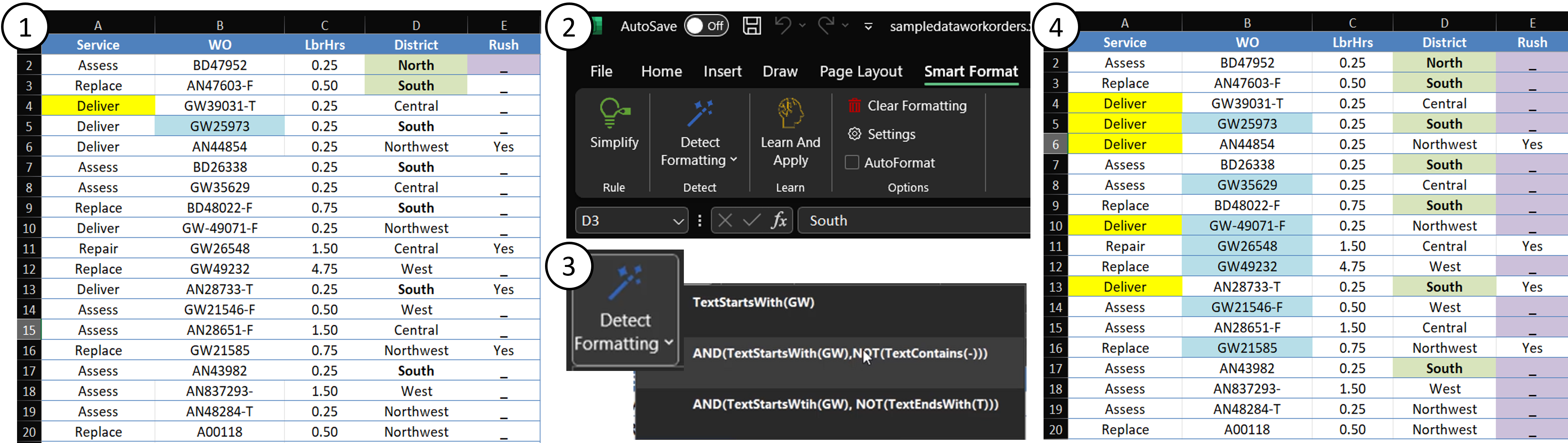} % Reduce the figure size so that it is slightly narrower than the column. Don't use precise values for figure width.This setup will avoid overfull boxes.
\caption{The \system{} user interface. \textbf{\circled{1}} Spreadsheet in Excel where the user has formatted a few cells as examples. \textbf{\circled{2}} \system{} UI in Excel ribbon as an Add-In listed under tab named \textit{Smart Format}. \textit{Learn and Apply} learns and applies the top rule on the selected column. \textbf{\circled{3}} \textit{Detect Formatting} lists the top 3 rules for the current column. \textbf{\circled{4}} The final sheet after applying the rules.}
\label{manual-demo}
\end{figure*}

\section{Solution Sketch}

A core challenge in \system{} is to learn
from few user-provided examples---often just one.
Traditional programming-by-example (PBE) systems~\cite{flashfill, flashextract} 
can carry out an efficient search 
because they can derive constraints
by relating the inputs and outputs 
for their task.
In conditional formatting, however, the output provides a relatively weak binary signal (whether a row is formatted or not).
The design of \system{} is meant to mitigate this challenge by \circled{1}
generating predicates that characterize
values in the target formatting column, \circled{2} hypothesizing labels for 
the entire column by clustering over the generated predicates, \circled{3} synthesizing many
rules that satisfy the hypothesized groups, and \circled{4} using a dedicated ranker that
furthers refines the likelihood of generated rules.
Figure~\ref{architecture} shows an overview of these steps, which we describe in following sections.

% Step \circled{1} enumerates properties of cells as predicates. Step \circled{2} approximates the expected output using semi-supervised clustering. \system{} then iteratively generates rules that match this output in step \circled{3}, and ranks them in step \circled{4}. 

\subsection{Predicate Generation}

% \jose{we are missing the first step: the user provides a column, with K formatted cells, then we go into this.}
% \jose{provide a bit of info on why we are generating predicates : to characterize values so we can then use this information in our clustering stage}
% The user provides a column with few formatted cells which \system{} then uses to learn the formatting rule.
\system{} computes properties of cell values in the form of boolean predicates---a boolean-valued function that takes a cell $c$, zero or more arguments and returns \textsf{true} if the property that it describes holds for $c$.
To avoid type errors, all predicates are assigned a type $t$ and they only match cells of their type. 
Mixed-type columns are assigned \emph{Text} type.
We chose predicates, shown in Table~\ref{tab:predicates}, based on the operations supported by popular spreadsheet platforms.

\begin{table}[t]
    \centering
    \caption{Supported predicates and their arguments for each data type. $c$ corresponds
    to a cell value,
    $n$ is a numeric literal,
    $d$ is a datetime literal,
    and $s$ is a string literal. \system{}
    uses type-specific
    generators to instantiate
    $n$, $d$, and $s$, as
    necessary.}
    \label{tab:predicates}
    \begin{tabularx}{.95\columnwidth}{lll}\toprule
      Numeric     & Datetime  & Text  \\ \midrule
      \textsf{greater($c$, $n$)} &\textsf{greater($c$, $n$, $d$)}& \textsf{equals($c$, $s$)}\\
      \textsf{greaterEquals($c$, $n$)}&\textsf{greaterEquals($c$, $n$, $d$)}&\textsf{contains($c$, $s$)}\\
      \textsf{less($c$, $n$)}&\textsf{less($c$, $n$, $d$)}&\textsf{startsWith($c$, $s$)}\\
      \textsf{lessEquals($c$, $n$)}&\textsf{lessEquals($c$, $n$, $d$)}&\textsf{endsWith($c$, $s$)}\\
      \textsf{between($c$, $n_1$, $n_2$)}&\textsf{between($c$, $n_1$, $n_2$, $d$)}&\\
    \bottomrule
    \end{tabularx}
\end{table}

% Predicate generation enumerates predicates that hold for a strict subset of the cells of the given column.
Given a column of cells and a predicate, the goal is to initialize each additional argument to a constant value such that the predicate returns \textsf{true} for a strict subset of the column.
We use different generators for each predicate type.
For example, \textsf{contains} uses tokens obtained by splitting on non-alphanumeric delimiters and also tokens from a prefix trie.
For numeric predicates, constants are generated from cell values, column statistics (mean, percentiles, min, max) and popular constants (0, 1, 100).
These predicates are then used for both clustering and constructing formatting rules.
We generate these predicates only for the column the user is currently formatting to prevent the predicate space from exploding.

\subsection{Semi-supervised Clustering}

Because the user has provided few examples, rule generation at this stage would only reflect the properties of these positive examples, but may not generalize to the full column.
To address this shortcoming, we first predict the expected output of the rule on the remainder of the unformatted cells.
We exploit positional information to constrain this problem: we assume users annotated their column top-down.
Any cells above (and between) formatted examples are intended to have no formatting,
while any cells after the last formatted example are true unassigned cells.
% We build this position information into a clustering algorithm.

% \jose{see if this informal description reads clearly enough}
We perform iterative clustering over our predicates to produce $k$ clusters, where $k$ is the number of unique formats in the column plus one cluster for unassigned cells.
% \jose{this idea of identifier comes out of the blue, either introduce before or just say unique formats} 
% For our column, 
Formatted examples and intentionally unformatted rows are never reassigned.
Unassigned cells are reassigned at each iteration based on the sum of minimum and maximum distance to any cluster element, where distance is computed as the symmetric difference between sets of predicates.
% (from our predicate generation step) that hold for each cell.
% Specifically, the cost
% between a cell and a cluster is defined as
% the sum of the minimal and maximal distance of the cell to every
% cell in that cluster, where we define distance to be the 
% size of the symmetric difference between the sets of predicates
% (from our predicate generation step) that hold for each cell.

% We define
% the distance between two cells is the size of the symmetric difference between the sets of predicates that hold for either cell.
% Taking inspiration from $k$-medoids \cite{kaufman2009finding} we iteratively reassign unformatted cells to a new cluster. Instead of computing a cluster medoid, however, we combine the minimal and maximal distance to any element of the cluster. This is computationally much more efficient (linear instead of quadratic in the number of distance computations) and was found to perform well in practice.
% When clusters become stable or a maximal number of iterations is reached, each cell takes the format value of their associated cluster.

\subsection{Candidate Rule Enumeration}

\system{} supports rules that can be built as a propositional formula in disjunctive normal form over predicates. In other words, every rule is of the form
$$\boldsymbol{\left(\right.} p_1(c) \wedge p_2(c) \wedge \ldots \boldsymbol{\left.\right)} \vee \boldsymbol{\left(\right.} p_j(c) \wedge p_{j+1}(c) \wedge \ldots\boldsymbol{\left.\right)} \vee \ldots$$
with $p_i$ a generated predicate or its negation. Our goal is to strike a balance between expressiveness and simplicity.

We greedily enumerate promising candidates by iteratively learning 
decision trees, which use our predicates as features to classify cells into the groupings produced by our semi-supervised clustering.
Each decision tree then corresponds to a rule in disjunctive normal form \cite{blockeel1998top}.
We identify and address three challenges: variety in rules, simplicity of rules, and coping with noisy labels.
To ensure variety, the root feature is removed from the set of candidate splits after each iteration.
To ensure simplicity, we only accept decision trees with $\lambda_n$ or fewer nodes ($\lambda_n$=10 as a default).
To deal with noisy labels, we only require decision trees to have perfect accuracy on user examples. In our greedy tree learning approach we consider labeled cells to be twice as important as unlabeled ones and we stop learning rules once accuracy falls below $\lambda_t$ ($\lambda_t=0.8$ as a default).

\subsection{Candidate Rule Ranking}

Because the rule learning step by construction produces multiple candidate rules, \system{} reorders candidates and returns the top rule to the user.
To perform this ranking, \system{} uses a neural ranker that considers, features of the data, the rule execution on that data~\cite{ranking_outputs} and intrinsic
properties of the rule~\cite{ranking_program_feats}, such as what predicates and literals
are used.
Information about the column is captured by turning the column data into a sequence of words and using a pre-trained language model~\cite{Devlin2019BERTPO} to obtain cell-level embeddings.
These embeddings are augmented with information about the execution of the rule through cross-attention \cite{CrossAttn}.
Information about the rule is captured by handpicked features.
We then concatenate the vectors of handpicked features and the vector capturing the column data and rule execution.
Finally, we learn a linear weight vector to reduce this feature vector to a single ranking score.
\footnote{We also designed a heuristic based symbolic ranker that performs slightly worse than the neural ranker. More details about the symbolic ranker are included in full paper.}
% We use this score, which captures both syntactic (rule properties) and
% semantic (data/execution) information, to 
% pick the most likely rule.

\section{Demonstration}
\begin{figure}[t]
\centering
\includegraphics[width=0.9\columnwidth]{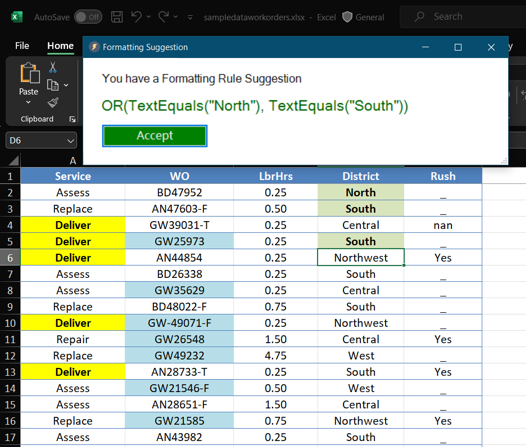} % Reduce the figure size so that it is slightly narrower than the column. Don't use precise values for figure width.This setup will avoid overfull boxes.
\caption{Interactive formatting rule suggestion by \system{}. The user edits the spreadsheet as usual and \system{} learns rules in the background, surfacing them after three examples have been provided.
% \jose{we never say when we are confident. If its a function of number of examples, let's be explicit. Otherwise we need to have a small section where we explain what we mean by confident and then reference here}. 
\system{} can help users who are unaware of conditional formatting rules in spreadsheet platforms.
% \jose{the formatting suggestion here looks like it is for Rush but the text describes one for District}\mukul{Fixed}
}
\label{interactive-demo}
\end{figure}

The participants can use \system{} on any workbook of their choice.
For this demonstration we use the Work Orders
% \jose{is this a known dataset? If not, then not sure we need a name, if it is, then briefly state used in prior work for XYZ, or state the source (e.g. public internet tutorial or whatever)} 
workbook from a public internet tutorial \cite{SampleWorkbook}.
The dataset contains information about work orders for a hypothetical services company. 
The data has 1000 rows and 22 attributes including order ID, district, number of technicians, service date and service type. 
% The dataset has 1000 rows out of which we take the first 100 rows as a reference dataset for this demonstration. 
% We expect that participants will be familiar with this data domain and will be able to understand the user's formatting intent and the rules suggested by \system{}\jose{I don't understand this sentence: why would the participants be familiar with the data domain? Should we just remove this?}.
We implemented a simple add-in for Microsoft Excel (called \emph{Smart Format}) to demonstrate \system{}.
Figures ~\ref{manual-demo} and~\ref{interactive-demo} show a screenshot of this add-in. 
% \jose{this sentence that follows seems wrong -- maybe this is from some prior demo paper used for guidance?}
% The top panel is for conformance constraint discovery and selection, and the bottom panel serves the purpose of interactive
% exploration of constraint violations by data tuples and data cleaning.

We will demonstrate two scenarios: a workflow where the user requests a rule from \system{} after providing a few formatting examples, and a workflow where \system{}, running in the background, can hand-raise and suggest a rule without prompting.
% First, we consider an experienced Excel user who employs \system{} to accelerate their regular workflow. 
% Second, we show how \system{} can help a novice user learn about and use CF rules for their formatting tasks. 
% First, we show 
% Second, we show 
The two workflows are designed to address different users' Excel expertise.
An experienced user performing sophisticated formatting tasks might prefer selecting rules themselves, exerting more
influence over \system{}.
A novice user, on the other hand, might be unaware of CF and benefit
from proactive suggestions for feature discovery.

% \jose{what is the key difference here? In one, we show a version of the workflow where the user requests a rule from \system{}. In the other, \system{}
% can raise its hand and suggest a rule 
% without prompting, by running in the background.
% Let's frame this way and then we can have a small discussion around why we expect experienced/novice users to benefit from each of these.
% }

\subsection{User Requested Suggestion}

For this demonstration, we will guide the participants through four steps.
We use Figure~\ref{manual-demo} to guide the demonstration. 

% \textbf{Step \circled{1} (Opening Reference Data):} The user opens the desired workbook in Excel. 
% For our guided scenario, the user opens the Work Orders dataset workbook as the reference data

\textbf{Step \circled{1} (Providing Examples):} The user colors a few cells as examples for each column that needs to be formatted.
For our scenario, the user wants to format ID, District, Service and Rush column, so they color one cell in each such column.
Their intent for each column is as follows:
% \jose{describe the intent for each of these. Maybe as a list?}
\begin{enumerate}
    \item \emph{Service} -- Fill yellow for all services that are delivery related.
    \item \emph{WO} -- IDs starting with \textit{``G''} or \textit{``AN''} and not ending with \textit{``T''}.
    \item \emph{District} -- Bold and fill green for North and South districts.
    \item \emph{Rush} -- Fill magenta for missing value cells denoted by \textit{"\_"}.
\end{enumerate}

\textbf{Step \circled{2} (Learning Formatting Rules):} To learn and apply rules, the user has 2 options. (1) The user can select which rule to apply by clicking on \textit{``Detect Formatting''} which opens a drop-down list of the top 3 rules that \system{} learned for the column that is currently active. 
% The user can choose the rule that fits their need. Hovering over a rule also displays a natural language description of the rule\jose{this comes out of nowhere, we need to describe it in the previous section or when we mention the add-in, state this is an extension to facilitate user interaction and how we generate these}. \mukul{I was not able to add this feature in anyway so removing}.
(2) The user can also choose to directly learn and apply the top ranked rule by selecting the desired columns and clicking \textit{``Learn and Apply''}.
% \jose{Confusing name, it has already learned, should we not call it apply?} 
% and \system{} will automatically apply the top ranked rule for each column in the selection. 
Option (2) allows for an in-context experience, as it can be triggered via a keyboard hotkey \texttt{(Ctrl+Shift+O)}.
% For our scenario, the user formats 1-2 cells based on the intent as described in step-\circled{2}.
% we select order ID, District, Service and Rush column and learn for this selection.\jose{You should remove this here and instead detail above what the intent for each column's formatting is.}

\textbf{Step \circled{4} (Updating Rules):} 
Like with other PBE systems, the user can improve the rules learned by 
\system{} by providing more examples.
For example, the user can give one example and learn the rule and if it is incorrect, the user can uncolor some cells or color new cells and relearn the rule. 
For this demonstration, lets consider the user wants to also color cells starting with \textit{"AN"} in the \emph{WO} column. The user colors the second cell (\textit{"AN47603-F"}) and then detects formatting rule. The user can now choose \texttt{AND(OR(Begins("GW"),Begins("AN")),NOT(EndsWith("T")))}.
\footnote{The user can also manually update the rule from the Conditional Formatting menu.}
% For this demonstration, the user wants to also color \textit{``Central''} values in the \textit{``District''} column. The user colors the cell \textit{``Central''} and then detects formatting rules. The user can now choose \texttt{OR(TextEquals("South"), TextEquals("Central"))}.\jose{Can we have a more interesting scenario from the video here?}

% \textbf{Step \circled{5} (Removing Rules):} Finally, we also allow easy removal of these rules. The user can select clear formatting to remove rules. 
% % For our scenario, the user selects the \textit{Rush} column and clears its formatting. \jose{This is not that interesting, so let's remove if we are tight on space.}

\subsection{Background Suggestion}
We now describe the hand-raise workflow of \system{} where the user edits the spreadsheet as normal and \system{} proactively suggests the top-ranked formatting rule to the user.
This workflow can help novice users who may be unfamiliar with \system{} or conditional formatting in general.
% \jose{reframe based on the prior suggestion: focus on workflow difference, not the user, and then say this workflow is particularly helpful for novices who may be unaware of CF etc etc}
% As mentioned in our introduction,
% unfortunately many novice users are not
% aware that conditional formatting exists.
% \system{} can help address the discoverability issues associated with conditional formatting.
For this demonstration, we will guide the participants through three steps
that exercise this proactive
workflow.
We use Figure~\ref{interactive-demo} to guide the demonstration. 

\textbf{Step \circled{1} (Opening Reference Data):} Similar to the first demonstration, the user opens the desired workbook in Excel. For this demonstration, we use the same Work Orders workbook as before.

\textbf{Step \circled{2} (Normal Editing):} 
The user edits the spreadsheet to format the \emph{District} column based on the intent described earlier.
% \jose{this step should just say the user normally starts... leave the learning in background to the next step.} 
In this scenario, the user wants to format \textit{``North''} and \textit{``South''} districts. The user starts editing the sheet and formats the first 2 cells in the district column that are either \textit{``North''} or \textit{``South''}.

\textbf{Step \circled{3} (Rule Suggestion):} \system{} can learn rules in the background as users are normally editing their spreadsheet. After three examples, \system{} is able to learn a rule in the background and can suggest this to the user, as shown in Figure~\ref{interactive-demo}, without any prompting from their end. Similarly to the prior scenario, the user can then decide to either accept the suggested rule, modify parameters, or give more examples to learn a better rule. The user accepts this rule and its application automatically formats the entire column as intended.
% \jose{same comment here regarding the more interesting scenario}

\bibliographystyle{ACM-Reference-Format}
\bibliography{references}

\end{document}